\documentclass[12pt,a4paper]{article}

\usepackage{amsbsy}
\usepackage{comment}
\usepackage{graphicx}
\usepackage{amsmath}
\usepackage{algorithm}
\usepackage[noend]{algpseudocode}

\newtheorem{definition}{Definition}

\begin{document}

\title{Using simulated annealing for locating array construction}
\author{{\normalsize
Tatsuya Konishi, Hideharu Kojima, Hiroyuki Nakagawa, Tatsuhiro Tsuchiya} \\
{\small Graduate School of Information Science and Technology, Osaka University} \\
{\small 1-5 Yamadaoka, Suita-City, Osaka 565-0871, Japan} \\ ~\\
{\small \textbf{Corresponding author} Tatsuhiro Tsuchiya (email: t-tutiya@ist.osaka-u.ac.jp)}
}
\date{}
\maketitle

\begin{abstract}
\noindent
\textbf{Context:} 
Combinatorial interaction testing is known to be an efficient testing strategy 
for computing and information systems. 
Locating arrays are mathematical objects that are useful for this testing strategy, 
as they can be used as a test suite that enables fault localization as well as fault detection.  
In this application, each row of an array is used as an individual test. 

~\\
\noindent
\textbf{Objective:} 
This paper proposes an algorithm for constructing locating arrays with a small number of rows.  
Testing cost increases as the number of tests increases;  
thus the problem of finding locating arrays of small sizes is of practical importance.   

~\\
\noindent
\textbf{Method:} 
The proposed algorithm uses simulation annealing, a meta-heuristic algorithm, to find locating array of a given size. 
The whole algorithm repeatedly executes the simulated annealing algorithm by 
dynamically varying the input array size.

~\\
\noindent
\textbf{Results:} 
Experimental results show 1) that the proposed algorithm is able to construct locating arrays for 
problem instances of large sizes and 2) that,  
for problem instances for which nontrivial locating arrays are known, 
the algorithm is often able to generate locating arrays that are 
smaller than or at least equal to the known arrays. 

~\\
\noindent
\textbf{Conclusion:} 
Based on the results, it is concluded that the proposed algorithm can produce small locating 
arrays and scale to practical problems. 
\end{abstract}


\section{Introduction}

This paper proposes to use simulated annealing to generate \emph{locating arrays}~\cite{colbourn_locatingarray2008}. 
A locating array can be used in \emph{combinatorial interaction testing}, 
a well-known strategy for testing computing and information systems~\cite{kuhn_introductionCT2013}. 
Combinatorial interaction testing aims to test all value combinations among some subsets of factors. 
The most common form of combinatorial interaction testing is $t$-way testing, 
which tests all value combinations among every set of $t$ factors.
The value of $t$ is referred to as \emph{strength} 
and a value combination among $t$ factors is said to be a $t$-way \emph{interaction} or  
an interaction of strength~$t$. 

As a running example, let us consider a simple model of a printer (Table~\ref{tab:sut}). 
The model consists of four factors: layout, size, color, and display.
The first three factors take two values, whereas the last factor takes three values. 
For example, the layout factor takes either portrait or  landscape.
A test for this model is a four-tuple where each element is a value of each factor;
thus there are a total of 24 possible tests. 

\begin{table}[b]
\centering
\caption{Printer example}\label{tab:sut}
\begin{tabular}{cccc}
	\hline 
	layout & size & color & duplex \\ 
	\hline \hline  
	portrait & A4 & Yes & oneside \\ 
	landscape & A5 & No & shortedge \\ 
	&  &  & longedge \\ 
	\hline 
\end{tabular} 
\end{table}

\emph{Covering arrays} are mathematical objects that play a key role in $t$-way testing.
A covering array of strength~$t$ can be regarded as 
a set of tests such that every $t$-way interaction appear at least once.  
Figure~\ref{fig:ca} shows a two-way covering array for this printer example. 
Each row represents a test. 
From this figure, it can be seen that 
every combination of values for two factors occurs in at least one of the six rows.
For example, among the size and duplex factors, a total of six 
interactions of strength two are possible, namely, (A4, oneside), (A4, shortedge), (A4, longedge), 
 (A5, oneside), (A5, shortedge), and (A5, longedge). 
Each one of these interactions appears somewhere in the array. 
Testing cost increases as the number of tests increases;  
thus the problem of finding a small covering arrays is of practical importance. 
Indeed, there has been a large body of research on generation of covering arrays of small size.



\begin{figure}[t]
\centering
\begin{tabular}{|c|c|c|c|c|} \hline
    & Layout   &  Size & Color & Duplex \\ \hline \hline
  1  & Portrait    & A4 & Yes & OneSide \\
  2  & Portrait    & A4 & No & ShortEdge  \\
  3  & Portrait    & A5 & No  & LongEdge   \\
  4  & Landscape    & A4 & Yes  & LongEdge    \\
  5  & Landscape   & A5 & Yes & ShortEdge \\
  6  & Landscape & A5 & No & OneSide  \\ \hline
\end{tabular}
\caption{Covering array for the printer example}
\label{fig:ca}
\end{figure}

\begin{figure}[t]\label{fig:la}
	\centering
	\begin{tabular}{|c|c|c|c|c|} \hline
		& Layout   &  Size & Color & Duplex \\ \hline \hline
		1  & Portrait    & A4 & Yes & OneSide \\
		2  & Portrait    & A4 & Yes & LongEdge  \\
		3  & Portrait    & A4 & No  & ShortEdge   \\
		4  & Portrait    & A5 & No  & OneSide    \\
		5  & Portrait    & A5 & No & LongEdge \\
		6  & Landscape & A4 & Yes & OneSide  \\
		7  & Landscape & A4 & Yes & ShortEdge \\
		8  & Landscape & A4 & No  & LongEdge  \\
		9  & Landscape & A5 & Yes  & OneSide  \\
		10  & Landscape & A5 & No & ShortEdge  \\ \hline
	\end{tabular}	
	\caption{Locating array for the printer example}
\end{figure}

\emph{Locating arrays}, which were proposed by 
Colbourn and McClary~\cite{colbourn_locatingarray2008}, 
add more values to covering arrays. 
For example, 
a $(\overline{d},  t)$-locating array enables to 
locate any set of interaction faults if there are at most $d$ faults and 
all the $d$ faults are of strength~$t$~\cite{colbourn_locatingarray2008}.  
Figure~\ref{fig:la} shows an example of a $(\overline{1}, 2)$-locating array for the printer model.
Now suppose that the outcome of executing a test is either \emph{pass} or \emph{fail}
and that the outcome is fail if and only if an interaction fault appears in the test. 
The array enables to locate any fault if it is a two-way interaction 
fault ($t=2$) and there is no other fault ($d = 1$).
For example, if the set of failing tests is $\{4, 5, 10\}$, then the faulty interaction 
is (A5, No), because no other two-way interaction yields this set of failing tests.

In spite of the useful ability described above, studies on locating arrays are 
still in an early stage and there are few ways of 
constructing locating arrays.  
The paper proposes a simulated annealing-based algorithm for generating $(\overline{1}, t)$-locating arrays. 
The algorithm uses this meta-heuristic algorithm for finding a locating array of a fixed size. 
By repeatedly running simulated annealing with different array sizes, the algorithm constructs 
a locating array of a small size. 
We conducted a set of experiments to evaluate the 
performance of the proposed algorithm. 
The results obtained show that 
the algorithm scales to problems of practical sizes 
when strength $t$ is two and 
that it is able to produce  
locating arrays of optimal or suboptimal sizes 
for small problem instances. 

This rest of the paper is organized as follows. 
Section~\ref{sec:locatingarray} presents the definition of locating arrays. 
Section~\ref{sec:sa} shows the simulated annealing algorithm for finding 
a locating array of a given size. 
Section~\ref{sec:algorithm} describes the whole algorithm which uses the 
simulated annealing algorithm as a subroutine. 
Section~\ref{sec:results} describes a set of experiments we have conducted 
and their results. 
Section~\ref{sec:realtedWork} summarizes related work.
Section~\ref{sec:conclusion} concludes the paper.

\section{Locating arrays}\label{sec:locatingarray}

The System Under Test (SUT) is modeled as $k$ integers $(v_1, v_2,\ldots,v_k)$ 
such that $v_i \geq 2$, where $v_i$ represents the number of values that can be 
taken by the $i$th factor. 
Unless otherwise explicitly mentioned, we mean by an \emph{array} $A = [a_{ij}]$ a matrix 
that has $k$ columns and satisfies the condition $a_{ij} \in \{0,...,v_j-1\}$. 
A matrix of no rows is also an array. 

An \emph{interaction} $T$ is a set of factor-value pairs each having a different factor. 
Formally, an interaction is a possibly empty subset of
$\{(j, V_j) : j\in\{1,...,k\}, V_j\in\{0,...,v_j-1\} \}$ 
such that no two distinct elements $(j, V), (j', V') \in T$ have the same factor, i.e., $j\neq j'$. 
An interaction $T$ is $t$-way or of strength~$t$ if and only if $|T| = t$ ($0\leq t \leq k$). 
An interaction of strength~0 is an empty set. 
We say that a row $(b_1, b_2,\ldots, b_k)$ of an array \emph{covers} an interaction 
$T$ if and only if $b_j = V$ for any $(j, V) \in T$.

Given an array $A$, we let $\rho_A(T)$ denote the set of rows that cover interaction $T$. 
Similarly, for a set $\cal T$ of interactions, we let 
$\rho_A({\cal T}) = \bigcup_{T\in {\cal T}} \rho_A(T)$.
Also we let ${\cal I}_t$ denote the set of interactions of strength $t$. 
For example, suppose that the SUT model is (2, 2, 2), which represents a system with three factors that have two values. 
Then ${\cal I}_2$ contains, for example, $\{(1, 0), (2, 1)\}$,  
$\{(1, 1), (3, 1)\}$, $\{(2, 0), (3, 0)\}$, etc.   



Colbourn and McClary introduce several versions of locating arrays.
Of them the following two are of our interest: 
\begin{definition}
	\label{def:LA1}
	An array $A$ is $(d, t)$-locating if and only if 
	\[
	\forall {\cal T}_1, {\cal T}_2 \subseteq {\cal I}_t 
	\mathrm{\ such\ that\ } |{\cal T}_1|=|{\cal T}_2| = d: 
	\rho_A({\cal T}_1) = \rho_A({\cal T}_2) 
	\Leftrightarrow 
	{\cal T}_1 = {\cal T}_2. 
	\]
\end{definition}

\begin{definition}
	\label{def:LA2}
	An array $A$ is $(\overline{d}, t)$-locating if and only if  
	\[
	\forall {\cal T}_1, {\cal T}_2 \subseteq {\cal I}_t 
	\mathrm{\ such\ that\ } |{\cal T}_1|\leq d, |{\cal T}_2| \leq d: 
	\rho_A({\cal T}_1) = \rho_A({\cal T}_2) 
	\Leftrightarrow 
	{\cal T}_1 = {\cal T}_2. 
	\]
\end{definition}


In these definitions, 
$\rho_A({\cal T}_1) = \rho_A({\cal T}_2) 
\Leftrightarrow 
{\cal T}_1 = {\cal T}_2$
is equivalent to 
\[
{\cal T}_1 \neq {\cal T}_2
\Rightarrow 
\rho_A({\cal T}_1) \neq \rho_A({\cal T}_2),  
\]
because 
${\cal T}_1 = {\cal T}_2 \Rightarrow 
\rho_A({\cal T}_1) = \rho_A({\cal T}_2)$
trivially holds for any ${\cal T}_1, {\cal T}_2$ by the definition of $\rho_A(\cdot)$. 
Hence, what effectively matters is only the other direction; i.e., 
${\cal T}_1 = {\cal T}_2 \Leftarrow \rho_A({\cal T}_1) = \rho_A({\cal T}_2)$. 

The intuition behind this definition is as follows. 
Let us assume that an interaction is either \emph{failure-triggering} 
or not and that the outcome of the execution of a test is 
\emph{fail} if the test covers at least one failure-triggering 
interaction; \emph{pass} otherwise. 
Given the test outcomes of all tests, 
a $(d, t)$-locating array or $(\overline{d}, t)$-locating array enables to identify any $d$ or any at most $d$ fault-triggering interactions 
of strength~$t$.  

\paragraph{Example} The following matrix is a mathematical representation of 
the $(\overline{1}, 2)$-locating array shown in Figure~\ref{fig:la}. 
The SUT is modeled as $(v_1, v_2, v_3, v_4) = (2, 2, 2, 3)$.
\[
\left[
\begin{array}{cccc}
0 & 0 & 0 & 0 \\
0 & 0 & 0 & 2 \\
0 & 0 & 1 & 1 \\ 
0 & 1 & 1 & 0 \\ 
0 & 1 & 1 & 2 \\ 
1 & 0 & 0 & 0 \\ 
1 & 0 & 0 & 1 \\ 
1 & 0 & 1 & 2 \\ 
1 & 1 & 0 & 0 \\ 
1 & 1 & 1 & 1 \\
\end{array}
\right]
\]
~\\

In the practice of software and systems testing, 
we are usually more interested in the case where the number of fault-triggering interactions
is at most one than in the case where the number of faults is 
exactly one or more than one. 
For this practical reason, we restrict ourselves to constructing  
$({\overline 1}, t)$-locating arrays.


Note that $\mathcal{I}_t = \emptyset$ if $|\mathcal{I}_t| = 0$. 
Hence the necessary and sufficient condition for an array $A$ to be $(\overline{1}, t)$-locating 
can be divided into two parts: one for the case where $\mathcal{T}_1 = \emptyset$ and $|\mathcal{T}_2| = 1$
and the other one for the case where $|\mathcal{T}_1| =  |\mathcal{T}_2| = 1$. 
That is, an array $A$ is $(\overline{1}, t)$-locating if and only if the two conditions hold: 
%
\begin{description}
	\item[Condition~1] $\forall T \in \mathcal{I}_t : \rho_A(T) \neq \emptyset$
	\item[Condition~2] $\forall T_1, T_2 \in \mathcal{I}_t : T_1 \neq T_2 \Rightarrow \rho_A(T_1) \neq \rho_A(T_2) $
\end{description}

\section{Simulated annealing}\label{sec:sa}

\subsection{Overview}
Simulated annealing is a class of meta-heuristic search algorithms. 
In our approach, simulated annealing is used to find a locating array 
of a given size. 
A small locating array is generated by repeatedly 
using simulated annealing while varying the array size. 
How to vary the size will be discussed in Section~\ref{sec:algorithm}. 

The pseudo-code in Figure~\ref{fig:annealing} shows the overview of our simulated annealing algorithm. 
The input of the algorithm is the array size, i.e., the number of rows $m$, 
and is given as the formal parameter of the function. 
The number of possible values for the factors, i.e., $(v_1, \ldots, v_k)$,  and  strength~$t$ 
are not explicitly described as the input, since they never change in the course of 
 repetition of the algorithm. 
The output is either a $(\overline{1}, t)$-locating array obtained 
or $\perp$ in which case the algorithm failed to 
find a locating array of size $m$. 

The algorithm starts with an $m\times k$ array with all entries randomly selected. 
The temperature $T$, which is a real value used for directing state search, is set to 
the initial value $t_{init}$.  
Then, an evolution step is repeated until a locating array is found or 
the number of repetitions reaches a predefined limit $k_{max}$. 

An evolution step begins with a random selection of another array $A'$ from the set of \emph{neighbors} of 
the current array $A$. 
A neighbor of an array $A$ is an array that is slightly different from $A$. 
The definition of neighbors and the way of selecting a neighbor can make significant effects on search performance. 

Cost function $f(\cdot)$ evaluates $A'$ with respect to how far it is from a locating array. 
Our proposed cost function, which will be described in Section~\ref{subsec:costfunction}, always takes a non-negative real value and reaches 0 when $A$ is $(\overline{1}, t)$-locating.  
The cost function directs the search, as it is used to determine whether or not to perform the change of the current array.  
If the change from $A$ to $A'$ does not increase the cost, i.e., $\Delta = f(A') - f(A) \leq 0$, then the change is always accepted. 
Otherwise, the change is performed with probability $\exp^{-\Delta/T}$, 
where $T$ is the temperature.
This probabilistic decision is intended to avoid trapping in local minima. 
Then, the temperature is decreased and the evolution step is repeated. 

\begin{figure}[t]
	\begin{algorithmic}[1]
        \State \textbf{input:} $m$ \Comment{Provided as the function's argument}
	    \State \textbf{output:} $(\overline{1},t)$-locating array $A$ or $\perp$
		\Function{SA}{$m$}
		\State Initialize $A$ \Comment{$A$ is an $m\times k$ array randomly generated}
		\State $T \gets t_{init}$ \Comment{$T$ represents the temparature.}
		\State $i \gets 0$
		\While{$i < k_{max}$}
			\State $A' \gets$ \textsc{SelectNeighbor}($A$)
			\Comment{Pick a random neighbor $A'$ of $A$}
            \If{$A'$ is $(\overline{1}, t)$-locating}
                \State \Return $A'$
            \EndIf
			\State $\Delta \gets f(A') - f(A)$
			\If{$\Delta \leq 0$}
				\State $A \gets A'$
			\Else
				\State $A \gets A'$ with probability $e^{-\Delta / T}$
			\EndIf
			\State $T \gets r * T$
				\Comment{$0 <  r < 1$}
			\State $i \gets i + 1$
		\EndWhile
		\State \Return $\perp$
	\EndFunction
	\end{algorithmic}
	\caption{Simulated annealing algorithm for finding a locating array of size $m$}
	\label{fig:annealing}
\end{figure}

\subsection{Cost function}\label{subsec:costfunction}
The cost function must be such that its value decreases if the 
current solution (array) becomes closer to a locating array. 
As stated in Section~\ref{sec:locatingarray}, the necessary and sufficient 
condition that an array is $(\overline{1}, t)$-locating consists of two parts. 
Based on this, we design a cost function as a sum of two functions, 
each evaluating each of the two parts.  
The cost function $f()$ is of the form:
\[ f(A) := weight * f_1(A) + f_2(A)
\]
where $f_1(A)$ and $f_2(A)$ correspond to Conditions~1 and~2, respectively, 
and $weight$ is a control parameter which takes a non-negative real value. 

We define:
\[ f_1(A) := |\{T \in \mathcal{I}_t : \rho_A(T) = \emptyset\}|
\]
In words, $f_1(A)$ is the number of interactions that are not covered by any row in $A$. 
The value of $f_1(A)$ reaches 0 if and only if all interactions are covered by at least 
one row in $A$. That is, Condition~1 is met if and only if $f_1(A) = 0$.

We define $f_2(T)$ as the number of $t$-way interactions that share 
identical covering rows with other interactions. 
Formally, 
\[
f_2(A) := \Big | \{ T \in \mathcal{I}_t : 
\exists T'\in \mathcal{I}_t [T\neq T' \land \rho_A(T)=\rho_A(T')\neq \emptyset] \} \Big |
\]
Clearly, Condition~2 is satisfied by $A$ if and only if the value of $f_2(A)$ is 0.
As a result, $A$ is  $(\overline{1}, t)$ locating if and only if $f(A) = 0$. 


\subsection{Neighbors and neighbor selection}\label{subsec:neighbors}

We consider two different strategies for selecting a neighbor of the current solution, 
i.e., array $A$. 
Each of the strategies depends on each definition of neighbors. 
One is straightforward (Figure~\ref{fig:neighborselection1}).
We regard $A' = [a'_{ij}]$ as a neighbor of $A = [a_{ij}]$ if and only if 
$A'$ is identical to $A$ except for one entry; i.e., 
$\Big|\{(a'_{i,j}, a_{i,j}) : a'_{i,j} \neq a_{i,j} \} \Big| = 1$. 
One neighbor is selected as the next solution from all such neighbors uniformly randomly.

The other strategy is intended to lead the search to an optimal solution more directly. 
In this strategy, 
an array $A'$ is regarded as a neighbor of the current array $A$ only if 
1) $\rho_{A'}(T) \neq \emptyset$ for some $T \in \mathcal{I}_t$ such that $\rho_A(T) = \emptyset$, 
or 2) $\rho_{A'}(T_1) \neq  \rho_{A'}(T_2)$
for some $T_1, T_2 \in \mathcal{I}_t$ such $\rho_A(T_1) = \rho_A(T_2)$.
The first condition states that $A'$ covers some interaction $T$ that is not covered by $A$.
The second condition signifies that for some two interactions $T_1$ and $T_2$, 
their covering rows are the same for $A$ but are different for $A'$. 
The purpose of restricting the candidates for the next array to these neighbors is to increase 
the likelihood that the next array will have a reduced cost. 

\begin{figure}[t]
	\begin{algorithmic}[1]
	    \State \textbf{input:} an $m\times k$ array $A$
	    \State \textbf{output:} an $m\times k$ array $A'$
		\Function{SelectNeighbor}{$A$}
		\State Randomly select an entry $a_{i, j}$ from $A$
		\State Randomly select $v$ from $\{0, 1, \ldots, v_j-1\} \backslash \{a_{i,j}\}$
		\State $A'$ $\leftarrow$ $A$ 
		with $a_{i,j}$ being replaced by $v$  
		\State \Return $A'$
	\EndFunction
	\end{algorithmic}
	\caption{Simple neighbor selection (baseline strategy)}
	\label{fig:neighborselection1}
\end{figure}

\begin{figure}[t]
	\begin{algorithmic}[1]
	    \State \textbf{input:} an $m\times k$ array $A$
	    \State \textbf{output:} an $m\times k$ array $A'$
		\Function{SelectNeighbor}{$A$}
		\State Let $I \leftarrow \{T \in \mathcal{I}_t : \rho_A(T) = \emptyset \}$  
		\If{$I \neq \emptyset$}
		  \State Randomly select an interaction $T$ from $I$
		  \State Randomly select a row $i$
		  \State $A'$ $\leftarrow$ 
		  $A$ with row $i$ being overwritten with $T$
		\Else
		    \State Randomly select an interaction $T$ 
		    \Statex \quad  \quad \quad  \quad
		    such that $\rho_A(T) = \rho_A(T')$ for another interaction $T'$
		    
		    \If{$|\rho_A(T)| > 1$ and a random Boolean value is true}
		    \State Randomly select a row $i$ such that $i \in \rho_A(T)$
		    \State Randomly select a factor $j$ involved in $T$ 
		    \State $A' \gets A$ with $a_{i,j}$ being altered
		    \Else
		    \State Randomly select a row $i$ such that $i \not\in \rho_A(T)$
		    \State $A' \gets A$ with row $i$ being 
		    overwritten with $T$
		    \EndIf
		\EndIf
		\State \Return $A'$
	\EndFunction
	\end{algorithmic}
	\caption{Neighbor selection (proposed strategy)}
	\label{fig:neighborselection2}
\end{figure}


The algorithm of this strategy proceeds as follows (Figure~\ref{fig:neighborselection2}):
First, if there is at least one interaction that is not covered by any row in $A$, 
the algorithm randomly picks one of such interactions, say $T$, and randomly selects one of the rows of $A$. Then it overwrites the row with the interaction $T$.


If all interactions of strength $t$ appear in the current solution $A$, 
randomly pick a $t$-way interaction $T$ such that 
the set of rows where it is covered is identical to the set of rows that cover
another $t$-way interaction $T'$. 
That is, $\rho_A(T) = \rho_A(T')$ holds. 
We want to have the next solution $A'$ such that $\rho_A'(T) \neq \rho_A'(T')$. 
To this end, one of the two ways is taken to perform randomly. 
One is to select a row that covers $T$ and
and randomly alter the value on one of the factors of $T$. 
The other one is to select a row that does not cover $T$ 
and overwrite it with $T$.

%
%

\section{Algorithm for constructing locating arrays}\label{sec:algorithm}



In this section, we present the whole algorithm for 
constructing locating arrays. 
This algorithm repeatedly executes simulated annealing presented in the previous section with array size $m$ being varied.

Binary search fits for the purpose of systematically changing the value of $m$. 
The input of the binary search is lower and upper bounds on the size of 
the smallest locating array, respectively. 
We let $low$ and $high$ denote the bounds. 
The first run of simulated annealing is executed for a size $\lfloor (low + high)/ 2 \rfloor$. 
If a locating array is found, then 
$high$ is updated to $\lfloor (low + high)/ 2 \rfloor - 1$; 
otherwise $low$ is updated to $\lfloor (low + high)/ 2 \rfloor + 1$. 
This process is repeated until $low > high$.  
Figure~\ref{fig:binarysearch} shows the binary search algorithm. 
The algorithm varies the array size and outputs 
the smallest locating array among the ones that can be obtained during the search.

As simulated annealing is probabilistic, the binary search may fail in finding a locating array even if there is indeed a locating array of size that falls between the given lower and upper bounds. 
In such a case, the output is $\perp$.

\begin{figure}[t]
	\begin{algorithmic}[1]
    \State \textbf{input:} integers $low$, $high$
    \State \textbf{output:} $(\overline{1},t)$-locating array $A$ or $\perp$
	\Function{Binary}{$low$, $high$}
		\State $A \gets \perp$
		\While{$low \leq high$}
			\State $size \gets \lfloor (low + high) /2 \rfloor$
			\State $A' \gets \textsc{SA}(size)$
			\If{$A' \neq \perp$}
				\State $A \gets A'$
				\State $high \gets size - 1$
			\Else
				\State $low \gets size + 1$
			\EndIf
		\EndWhile
		\State \Return $A$
		\EndFunction
		\end{algorithmic}
	\caption{Binary search algorithm using simulated annealing. }
\label{fig:binarysearch}
\end{figure}


Similarly, each run of simulated annealing can fail in obtaining a locating array of a given size. 
Having these in mind, we propose a two-phase algorithm as shown in Figure~\ref{fig:wholealgorithm}. 
The first phase is a fall-back of the binary search. 
In this phase, binary search is repeated until a locating array  is obtained. 
The second phase is used to improve the solution that has already been obtained. 
In this phase, simulated annealing is repeated at most $max$ times for the size smaller by one than the array 
obtained in the first phase. 
If a locating array is obtained during the $max$ runs, the size is further decreased by one and the phase is repeated. 
The termination of the algorithm is ensured by timeout. 
If timeout occurs, the algorithm will output the smallest locating array found 
during its execution. The output is $\perp$ if no locating array has been obtained. 

One problem that remains to be solved is how to compute 
the lower and upper bounds (Line~4 in Figure~\ref{fig:wholealgorithm}). 
At this moment, we obtain these bounds using the lower 
bound formula provided by Tang, Colbourn, and Yin~\cite{tang_optimalitylocatingarray_2012}. 
The formula is: 
\[
\min \left\{
\left \lceil \frac{2 \tbinom{k}{t}  v^t}{1+ \tbinom{k}{t}} \right\rceil, 
\left\lceil  -\frac{3}{2} - \dbinom{k}{t} + \sqrt{\dbinom{k}{t}^2 + (3+6v^t)\dbinom{k}{t} + \frac{9}{4}} \right\rceil
\right\}
\]
This lower bound formula assumes that the number of values that a factor can take 
is the same for all factors; i.e., all $v_i$ are the same $v$. 
Although in our case, $v_i$ can be different for different $i$, 
this formula can be used to obtain rough bounds as follows. 
We select the smallest $v_i$, denoted as $\min\{v_i\}$, and then apply that value to the formula to obtain a lower bound. 
To obtain an upper bound, we select the greatest $v_i$, 
denoted as $\max\{v_i\}$, and apply $\max\{v_i\}+1$ to the formula. 
Strictly speaking, the upper bound obtained this way is merely 
an approximation in the sense that there is no mathematical 
guarantee that the value is actually an upper bound. 
In practice, however, it has turned out that 
our approximation indeed serves as upper bounds for all problem instances tested 
in the experiments which will be described in the next section.    


\begin{figure}[t]
	\begin{algorithmic}[1]
    \State \textbf{output:} $(\overline{1},t)$-locating array $A$ or $\perp$
    \Statex
    \State \textbf{main algorithm:}
	\State $A \gets \perp$
	\State Compute the $lower$ and $upper$ bounds on the array size 
    \While{$A = \perp$}   \Comment{Phase 1}
      \State $A \gets$ \textsc{Binary}$(lower, upper)$
  	\EndWhile
	\State $i \leftarrow 0$, $size \gets size(A) - 1$  
	\While{$i < max$ and $lower \leq size$}\Comment{Phase 2}
		\State{$A' \gets \textsc{SA}(size)$}
		\If{$A' \neq \perp$}
			\State{$A \gets A'$}
			\State{$i \gets 0$, $size \gets size - 1$}
		\Else
			\State{{$i \leftarrow i + 1$}}
		\EndIf
	\EndWhile
	\State Output $A$ and terminate 
	\Statex
	\State \textbf{timeout:}
	    \State Output $A$ and terminate 
	\end{algorithmic}
	\caption{The whole algorithm. The algorithm repeats binary search 
	until a locating array is obtained (Phase~1) and then gradually decreases 
	the input array size (Phase~2).}
	\label{fig:wholealgorithm}
\end{figure}

%
%
%

\section{Experimental results}\label{sec:results}

We developed a program  that implements the proposed algorithm 
using the Java language. 
This section presents the results of experiments we conducted 
using the program. 
All experiments were performed on a Windows~10 machine equipped with 
16 GB memory and a 2.0 GHz Xeon E6540 CPU. 
The program and problem instances used in the experiments 
are available at Github at 
\texttt{https://github.com/tatsuhirotsuchiya/SA4LA}.

\subsection{Experiment 1}
We conduct three sets of experiments. 
The purpose of the first set is two-fold. 
First, it is intended to compare the two neighbor selection strategies 
of simulated annealing. As stated in Section~\ref{subsec:neighbors}, we 
consider two different approaches. The first one selects an array that is different from the current solution 
in a single entry as the next solution. 
The second one chooses the next solution by ``overwriting'' the current array 
with an interaction so that the search can fast converge to a locating array. 
Thereafter, we call the first strategy \textit{baseline} and the second one the 
\textit{proposed strategy}. 
The second purpose is to determine the values of control parameters  
for the remaining experiments. 
While performing the comparison of the two neighbor selection strategies, 
we tested a number of different settings of control parameters.
Hence the results can also be used for the second purpose. 

To this end, we selected one problem instance, called spin-s, from a well-known benchmark for 
combinatorial interaction testing research~\cite{garvin_metaheuristic2011,lin_TCA2015,yamada_satsolving2015}. 
This problem instance is the smallest among the 35 instances in the benchmark. 
Note that this choice favors the baseline strategy, because the strategy searches the solution space 
in finer steps than the proposed one.  
We considered the case $t = 2$; that is, we considered the problem of constructing $(\overline{1},2)$-locating arrays. 

\begin{figure}[t]
    \centering
    \includegraphics[scale=0.4]{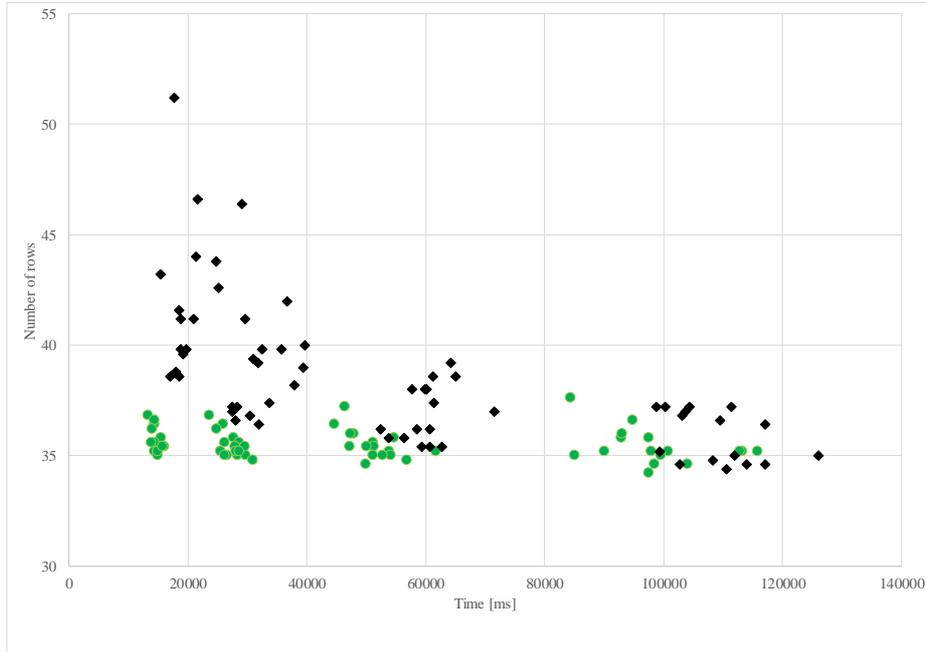}
    \caption{Size (number of rows) and computation time for different parameter settings (problem name: spin-s). 
    The diamonds and circles represent the results of the baseline strategy and the proposed strategy, respectively.}
    \label{fig:graph1}
\end{figure}

We tested a total of 64 parameter settings for each strategy. 
Specifically we tested $weight \in \{0.5, 1.0, 2.0, 4.0\}$, 
$t_{init} \in \{0.5, 1.0, 2.0, 4.0\}$, $k_{max} \in \{2048$, $4096$, $8192$, $16384\}$. 
The values of $r$ (the temperature cooling rate in Figure~\ref{fig:annealing})  
and $max$ (the maximum number of repetitions of Phase~2 in Figure~\ref{fig:wholealgorithm})  were set as follows: $r = 0.999$, and $max = 3$. 
For each of the settings, we ran the program five times and measured the average 
size of the resulting locating arrays and the average computation time. 
Figure~\ref{fig:graph1} summarizes the results. Different dot types represent different strategies. 
From the results, it can be clearly seen that the proposed strategy outperformed the baseline strategy.
Therefore, we only consider the proposed strategy in the remaining sets of experiments. 

The circles represent the results of the proposed strategy. 
They were widely distributed horizontally but fairly concentrated vertically. 
This means that computation time, which highly depends on the value of 
$k_{max}$, does not make significant contribution to reducing the array size. 
Hence for the remaining part of experiments, we set $k_{max}$ to 2048, the smallest value 
of the four values we tested. 
Other parameter values were taken from the case when the smallest size had been obtained 
when $k_{max} = 2048$; namely, $weight = 4.0$ and $t_{init} = 0.5$. 

\begin{table}[p]
\centering
\caption{Performance in generating $(\overline{1}, 2)$-locating arrays. 
The algorithm was repeated five times for each problem instance. 
The unit of time is second.}\label{tbl:result2}

\small
\begin{tabular}{lllllllll}
\hline 
 & & & \multicolumn{3}{c}{$x$ runs} & \multicolumn{3}{c}{$y$ runs} \\
 name    & factors                        & $x$/$y$/5 & time   & rows & min & time   & rows  & min \\ \hline 
apache   & $2^{158}3^{8}4^{4}5^{1}6^{1}$  & 0/5/5 & -      & -    & -   & 2062.5 & 64.2  & 63  \\
gcc      & $2^{189}3^{10}$                & 0/5/5 & -      & -    & -   & 1581.3 & 39.8  & 39  \\
bugzilla & $2^{49}3^{1}4^{2}$             & 5/5/5 & 167.0  & 32.0 & 32  & 82.4   & 32.0  & 32  \\
spin-s    & $2^{13}4^{5}$                  & 5/5/5 & 39.2   & 34.4 & 34  & 24.6   & 34.4  & 34  \\
spin-v    & $2^{42}3^{2}4^{11}$            & 5/5/5 & 393.9  & 50.2 & 50  & 234.5  & 50.2  & 50  \\
1        & $2^{86}3^{3}4^{1}5^{5}6^{2}$   & 5/5/5 & 2274.1 & 74.4 & 74  & 1199.5 & 74.4  & 74  \\
2        & $2^{86}3^{3}4^{3}5^{1}6^{1}$   & 5/5/5 & 1844.2 & 56.0 & 55  & 1118.0 & 56.0  & 55  \\
3        & $2^{27}4^{2}$                  & 5/5/5 & 43.7   & 28.0 & 28  & 11.0   & 28.0  & 28  \\
4        & $2^{51}3^{4}4^{2}5^{1}$        & 5/5/5 & 342.9  & 41.2 & 41  & 161.1  & 41.2  & 41  \\
5        & $2^{155}3^{7}4^{3}5^{5}6^{4}$  & 0/5/5 & -      & -    & -   & 3212.3 & 92.2  & 89  \\
6        & $2^{73}4^{3}6^{1}$             & 5/5/5 & 735.4  & 49.8 & 49  & 363.0  & 49.8  & 49  \\
7        & $2^{29}3^{1}$                  & 5/5/5 & 34.4   & 21.0 & 21  & 11.0   & 21.0  & 21  \\
8        & $2^{109}3^{2}4^{2}5^{3}6^{3}$  & 2/5/5 & 3088.2 & 78.5 & 77  & 1701.8 & 78.2  & 77  \\
9        & $2^{57}3^{1}4^{1}5^{1}6^{1}$   & 5/5/5 & 408.3  & 48.2 & 48  & 234.5  & 48.2  & 48  \\
10       & $2^{130}3^{6}4^{5}5^{2}6^{4}$  & 0/5/5 & -      & -    & -   & 2786.0 & 85.8  & 84  \\
11       & $2^{84}3^{4}4^{2}5^{2}6^{4}$   & 5/5/5 & 2585.4 & 80.2 & 79  & 1670.2 & 80.2  & 79  \\
12       & $2^{136}3^{4}4^{3}5^{1}6^{3}$  & 0/5/5 & -      & -    & -   & 2533.7 & 78.0  & 77  \\
13       & $2^{124}3^{4}4^{1}5^{2}6^{2}$  & 0/5/5 & -      & -    & -   & 3072.8 & 69.0  & 68  \\
14       & $2^{81}3^{5}4^{3}6^{3}$        & 5/5/5 & 1751.7 & 71.0 & 70  & 681.1  & 71.0  & 70  \\
15       & $2^{50}3^{4}4^{1}5^{2}6^{1}$   & 5/5/5 & 502.9  & 55.4 & 54  & 308.4  & 55.4  & 54  \\
16       & $2^{81}3^{3}4^{2}6^{1}$        & 5/5/5 & 1179.2 & 51.2 & 50  & 600.7  & 51.2  & 50  \\
17       & $2^{128}3^{3}4^{2}5^{1}6^{3}$  & 0/5/5 & -      & -    & -   & 2780.7 & 75.6  & 75  \\
18       & $2^{127}3^{2}4^{4}5^{6}6^{2}$  & 0/5/5 & -      & -    & -   & 2363.5 & 81.2  & 81  \\
19       & $2^{172}3^{9}4^{9}5^{3}6^{4}$  & 0/5/5 & -      & -    & -   & 2289.5 & 97.0  & 97  \\
20       & $2^{138}3^{4}4^{5}5^{4}6^{7}$  & 0/5/5 & -      & -    & -   & 2987.9 & 105.2 & 102 \\
21       & $2^{76}3^{3}4^{2}5^{1}6^{3}$   & 5/5/5 & 1965.6 & 71.2 & 71  & 1346.7 & 71.2  & 71  \\
22       & $2^{73}3^{3}4^{3}$             & 5/5/5 & 1262.4 & 55.0 & 54  & 840.8  & 55.0  & 54  \\
23       & $2^{25}3^{1}6^{1}$             & 5/5/5 & 72.1   & 36.0 & 36  & 45.2   & 36.0  & 36  \\
24       & $2^{110}3^{2}5^{3}6^{4}$       & 2/5/5 & 3366.5 & 81.5 & 81  & 2216.3 & 81.6  & 81  \\
25       & $2^{118}3^{6}4^{2}5^{2}6^{6}$  & 0/5/5 & -      & -    & -   & 2868.5 & 92.8  & 92  \\
26       & $2^{87}3^{1}4^{3}5^{4}$        & 5/5/5 & 1509.7 & 59.6 & 59  & 817.2  & 59.6  & 59  \\
27       & $2^{55}3^{2}4^{2}5^{1}6^{2}$   & 5/5/5 & 571.5  & 60.6 & 60  & 340.3  & 60.6  & 60  \\
28       & $2^{167}3^{16}4^{2}5^{3}6^{6}$ & 0/5/5 & -      & -    & -   & 2066.3 & 110.6 & 97  \\
29       & $2^{134}3^{7}5^{3}$            & 0/5/5 & -      & -    & -   & 2668.2 & 58.2  & 57  \\
30       & $2^{72}3^{4}4^{1}6^{2}$        & 5/5/5 & 985.6  & 38.4 & 37  & 680.4  & 38.4  & 37 \\ \hline 
\end{tabular}

\end{table}

\subsection{Experiment 2}
The purpose of this set of experiments is to examine if the algorithm scales to 
large problems that arise in practice.  
To this end, we applied the proposed algorithm to 35 problem instances presented in \cite{garvin_metaheuristic2011}. 
This collection of problem instances has been used 
in several studies in combinatorial interaction testing 
(e.g.,~\cite{garvin_metaheuristic2011}).
Table~\ref{tbl:result2} lists these instances. 
The first column shows the problem name. 
In the second column, $f_1^{l_1} f_2^{l_2} \ldots $ represents that 
the problem instance (SUT model) has 
$l_i$ factors that take $f_i$ values ($i = 1, 2, \ldots$).

As in the first set of experiments, 
we ran the algorithm for each instance five times. 
We set timeout period to one hour for each run. 
Compared to the problem instance used in the first experiment, many of the instances were large in size,  
so that the algorithm often failed to complete its execution within the timeout period. 
Nevertheless, it was often the case that a locating array was obtained, in which case 
at least one locating array was constructed during algorithm execution. 

Having this in mind, we summarize the results obtained in Table~\ref{tbl:result2} for 
the case $t = 2$ and Table~\ref{tbl:result3} for the case $t = 3$.  
Each row corresponds to each of the 35 problem instances. 
For each problem instance, we show two numbers, $x$ and $y$, in the form of $x/y/5$. 
Here $x$ is the number of runs that were completed within the timeout period, while $y$ is 
the number of runs in which at least one $(\overline{1}, t)$-locating array was obtained. 
Hence $0\leq x \leq y \leq 5$ always hold. 

The forth and five columns from left show 
the total running time and size of the resulting locating array averaged over the $x$ finished runs.
The sixth column labeled ``min'' shows the size of the smallest array among these $x$ locating arrays.

Similarly, the time spent from the start until the last locating array was found 
and the size of that array are averaged over the $y$ runs and shown in the third and second 
columns from right. 
(Note that the algorithm needs to continue to run after 
the last locating array obtained.)
The rightmost column shows the size of the smallest array obtained in the $y$ runs. 

The results shown in Table~\ref{tbl:result2} shows that 
the proposed algorithm scales well to these 
large problems when the strength $t$ is two. 
To our knowledge, no previous studies report 
locating arrays of strength two that  
are as large (in terms of the number of columns) as those obtained in this set of experiments. 

When strength $t$ is three (Table~\ref{tbl:result3}), our program failed 
to generate a locating array for many of the 
problem instances, mainly because 
the program ran out of memory. 
This seems to suggest that further improvement in algorithm design
or implementation is required. 
Still, the program was able to 
produce locating arrays for 11 instances. 
To the best of our knowledge, there are no previous studies 
that reported locating arrays of strength three whose sizes 
are comparable to those we obtained in this set of experiments.

\begin{table}[p]
\centering
\caption{Performance in generating $(\overline{1}, 3)$-locating arrays}\label{tbl:result3}
\small

\begin{tabular}{lllllllll}
\hline 
 & & & \multicolumn{3}{c}{$x$ runs} & \multicolumn{3}{c}{$y$ runs} \\
 name    & factors                        & $x$/$y$/5 & time   & rows & min & time   & rows  & min \\ \hline 
apache   & $2^{158}3^{8}4^{4}5^{1}6^{1}$  & 0/0/5 & -      & -     & -   & -      & -     & -   \\
gcc      & $2^{189}3^{10}$                & 0/0/5 & -      & -     & -   & -      & -     & -   \\
bugzilla & $2^{49}3^{1}4^{2}$             & 0/5/5 & -      & -     & -   & 2858.4 & 150.0 & 150 \\
spin-s    & $2^{13}4^{5}$                  & 5/5/5 & 1247.5 & 159.6 & 158 & 896.8  & 159.6 & 158 \\
spin-v    & $2^{42}3^{2}4^{11}$            & 0/0/5 & -      & -     & -   & -      & -     & -   \\
1        & $2^{86}3^{3}4^{1}5^{5}6^{2}$   & 0/0/5 & -      & -     & -   & -      & -     & -   \\
2        & $2^{86}3^{3}4^{3}5^{1}6^{1}$   & 0/0/5 & -      & -     & -   & -      & -     & -   \\
3        & $2^{27}4^{2}$                  & 4/5/5 & 2732.2 & 107.0 & 106 & 1753.4 & 107.0 & 106 \\
4        & $2^{51}3^{4}4^{2}5^{1}$        & 0/5/5 & -      & -     & -   & 97.1   & 329.0 & 329 \\
5        & $2^{155}3^{7}4^{3}5^{5}6^{4}$  & 0/0/5 & -      & -     & -   & -      & -     & -   \\
6        & $2^{73}4^{3}6^{1}$             & 0/5/5 & -      & -     & -   & 1079.9 & 351.0 & 351 \\
7        & $2^{29}3^{1}$                  & 5/5/5 & 1945.5 & 62.2  & 61  & 1120.5 & 62.2  & 61  \\
8        & $2^{109}3^{2}4^{2}5^{3}6^{3}$  & 0/0/5 & -      & -     & -   & -      & -     & -   \\
9        & $2^{57}3^{1}4^{1}5^{1}6^{1}$   & 0/5/5 & -      & -     & -   & 93.6   & 516.0 & 516 \\
10       & $2^{130}3^{6}4^{5}5^{2}6^{4}$  & 0/0/5 & -      & -     & -   & -      & -     & -   \\
11       & $2^{84}3^{4}4^{2}5^{2}6^{4}$   & 0/0/5 & -      & -     & -   & -      & -     & -   \\
12       & $2^{136}3^{4}4^{3}5^{1}6^{3}$  & 0/0/5 & -      & -     & -   & -      & -     & -   \\
13       & $2^{124}3^{4}4^{1}5^{2}6^{2}$  & 0/0/5 & -      & -     & -   & -      & -     & -   \\
14       & $2^{81}3^{5}4^{3}6^{3}$        & 0/0/5 & -      & -     & -   & -      & -     & -   \\
15       & $2^{50}3^{4}4^{1}5^{2}6^{1}$   & 0/5/5 & -      & -     & -   & 185.0  & 515.0 & 515 \\
16       & $2^{81}3^{3}4^{2}6^{1}$        & 0/5/5 & -      & -     & -   & 1229.2 & 351.0 & 351 \\
17       & $2^{128}3^{3}4^{2}5^{1}6^{3}$  & 0/0/5 & -      & -     & -   & -      & -     & -   \\
18       & $2^{127}3^{2}4^{4}5^{6}6^{2}$  & 0/0/5 & -      & -     & -   & -      & -     & -   \\
19       & $2^{172}3^{9}4^{9}5^{3}6^{4}$  & 0/0/5 & -      & -     & -   & -      & -     & -   \\
20       & $2^{138}3^{4}4^{5}5^{4}6^{7}$  & 0/0/5 & -      & -     & -   & -      & -     & -   \\
21       & $2^{76}3^{3}4^{2}5^{1}6^{3}$   & 0/0/5 & -      & -     & -   & -      & -     & -   \\
22       & $2^{73}3^{3}4^{3}$             & 0/0/5 & -      & -     & -   & -      & -     & -   \\
23       & $2^{25}3^{1}6^{1}$             & 5/5/5 & 3278.5 & 132.0 & 131 & 2203.0 & 132.0 & 131 \\
24       & $2^{110}3^{2}5^{3}6^{4}$       & 0/0/5 & -      & -     & -   & -      & -     & -   \\
25       & $2^{118}3^{6}4^{2}5^{2}6^{6}$  & 0/0/5 & -      & -     & -   & -      & -     & -   \\
26       & $2^{87}3^{1}4^{3}5^{4}$        & 0/0/5 & -      & -     & -   & -      & -     & -   \\
27       & $2^{55}3^{2}4^{2}5^{1}6^{2}$   & 0/5/5 & -      & -     & -   & 644.2  & 516.0 & 516 \\
28       & $2^{167}3^{16}4^{2}5^{3}6^{6}$ & 0/0/5 & -      & -     & -   & -      & -     & -   \\
29       & $2^{134}3^{7}5^{3}$            & 0/0/5 & -      & -     & -   & -      & -     & -   \\
30       & $2^{72}3^{4}4^{1}6^{2}$        & 0/0/5 & -      & -     & -   & -      & -     & -  \\ \hline
\end{tabular}

\end{table}

\subsection{Experiment 3}

\begin{table}[p]
    \centering
    \caption{Results for problem instances for which nontrivial 
    $(\overline{1}, 2)$-locating arrays were already known.}
    \small
    \label{tab:known}
    \begin{tabular}{lllllllll}
    \hline 
  & known & &  \multicolumn{3}{c}{$x$ runs} & \multicolumn{3}{c}{$y$ runs} \\
 factors    &  size                    & $x$/$y$/5 & time   & rows & min & time   & rows  & min \\ \hline 
\begin{tabular}[c]{@{}l@{}}$2^{28} 3^9 4^6 5^4 6^{10}$ \\ 
$7^5 8^4 9^1 10^8$\end{tabular} & 421~\cite{Aldaco:2015}        & 0/5/5 & -     & -    & -   & 3055.7 & 333.0 & \textbf{333} \\
$2^3 3^7 4^5 5^9$                                                                        & 109~\cite{Compton2016}        & 5/5/5 & 168.0 & 74.4 & 74  & 113.9  & 74.4  & \textbf{74}  \\
$2^3$                                                                                    & \textbf{\underline{6}}~\cite{tang_optimalitylocatingarray_2012}          & 5/5/5 & 0.1   & 6.0  & 6   & 0.1    & 6.0   & \textbf{\underline{6}}   \\
$2^4$                                                                                    & \textbf{\underline{7}}~\cite{konishi_LASAT2017}          & 5/5/5 & 0.1   & 7.0  & 7   & 0.1    & 7.0   & \textbf{\underline{7}}   \\
$2^5$                                                                                    & \textbf{\underline{8}}~\cite{konishi_LASAT2017}          & 5/5/5 & 0.1   & 8.0  & 8   & 0.1    & 8.0   & \textbf{\underline{8}}   \\
$2^6$                                                                                    & \textbf{\underline{9}}~\cite{konishi_LASAT2017}          & 5/5/5 & 0.4   & 9.0  & 9   & 0.4    & 9.0   & \textbf{\underline{9}}   \\
$2^7$                                                                                    & \textbf{\underline{10}}~\cite{konishi_LASAT2017}         & 5/5/5 & 0.2   & 10.0 & 10  & 0.2    & 10.0  & \textbf{\underline{10}}  \\
$2^8$                                                                                    & \textbf{\underline{11}}~\cite{konishi_LASAT2017}         & 5/5/5 & 2.8   & 11.0 & 11  & 0.2    & 11.0  & \textbf{\underline{11}}  \\
$2^9$                                                                                    & \textbf{\underline{11}}~\cite{konishi_LASAT2017}         & 5/5/5 & 3.2   & 11.6 & 11  & 1.4    & 11.6  & \textbf{\underline{11}}  \\
$2^{10}$                                                                                 & \textbf{\underline{11}}~\cite{konishi_LASAT2017}         & 5/5/5 & 2.8   & 11.6 & 11  & 1.5    & 11.6  & \textbf{\underline{11}}  \\
$2^{11}$                                                                                 & \textbf{\underline{11}}~\cite{konishi_LASAT2017}         & 5/5/5 & 5.4   & 13.0 & 13  & 1.6    & 13.0  & 13  \\
$2^{12}$                                                                                 & \textbf{\underline{12}}~\cite{konishi_LASAT2017}         & 5/5/5 & 7.5   & 13.6 & 13  & 2.0    & 13.6  & 13  \\
$2^{13}$                                                                                 & \textbf{14}~\cite{konishi_LASAT2017}         & 5/5/5 & 6.9   & 14.0 & 14  & 2.3    & 14.0  & \textbf{14}\\
$2^{14}$                                                                                 & 15~\cite{konishi_LASAT2017}         & 5/5/5 & 10.3  & 14.4 & 14  & 4.1    & 14.4  & \textbf{14}  \\
$2^{15}$                                                                                 & \textbf{15}~\cite{konishi_LASAT2017}         & 5/5/5 & 10.8  & 15.0 & 15  & 3.2    & 15.0  & \textbf{15}  \\
$2^{16}$                                                                                 & \textbf{15}~\cite{Konishi2019arXiv}         & 5/5/5 & 12.1  & 15.0 & 15  & 3.9    & 15.0  & \textbf{15}  \\
$2^{17}$                                                                                 & \textbf{16}~\cite{Konishi2019arXiv}          & 5/5/5 & 11.4  & 16.0 & 16  & 3.5    & 16.0  & \textbf{16}  \\
$2^{18}$                                                                                 & \textbf{16}~\cite{Konishi2019arXiv}          & 5/5/5 & 14.2  & 16.0 & 16  & 6.4    & 16.0  & \textbf{16}  \\
$2^{19}$                                                                                 & 17~\cite{Konishi2019arXiv}          & 5/5/5 & 15.3  & 16.4 & 16  & 5.8    & 16.4  & \textbf{16}  \\
$2^{20}$                                                                                 & \textbf{17}~\cite{Konishi2019arXiv}          & 5/5/5 & 16.9  & 17.0 & 17  & 5.4    & 17.0  & \textbf{17}  \\
$2^{21}$   & 18~\cite{Konishi2019arXiv}          & 5/5/5 & 19.2  & 17.0 & 17  & 5.8    & 17.0  & \textbf{17}  \\
$2^{22}$  & 18~\cite{Konishi2019arXiv}          & 5/5/5 & 24.5  & 17.4 & 17  & 12.8   & 17.4  & \textbf{17}  \\
$2^{23}$                               & 19~\cite{Konishi2019arXiv}          & 5/5/5 & 23.2  & 18.0 & 18  & 10.5   & 18.0  & \textbf{18}  \\
$3^3$                                                                                    & \textbf{\underline{15}}~\cite{konishi_LASAT2017}         & 5/5/5 & 0.8   & 15.0 & 15  & 0.1    & 15.0  & \textbf{\underline{15}}  \\
$3^4$                                                                                    & \textbf{\underline{16}}~\cite{konishi_LASAT2017}         & 5/5/5 & 1.1   & 16.0 & 16  & 1.1    & 16.0  & \textbf{\underline{16}}  \\
$3^5$                                                                                    & \textbf{\underline{17}}~\cite{konishi_LASAT2017}         & 5/5/5 & 2.8   & 18.6 & 18  & 1.2    & 18.6  & 18  \\
$3^6$                                                                                    & \textbf{\underline{17}}~\cite{konishi_LASAT2017}         & 5/5/5 & 4.1   & 20.6 & 20  & 2.0    & 20.6  & 20  \\
$3^7$                                                                                    & 23~\cite{konishi_LASAT2017}         & 5/5/5 & 5.2   & 22.6 & 22  & 2.0    & 22.6  & \textbf{22}  \\
$3^8$                                                                                    & 25~\cite{konishi_LASAT2017}         & 5/5/5 & 7.1   & 24.0 & 24  & 3.3    & 24.0  & \textbf{24}  \\
$3^9$                                                                                    & 27~\cite{konishi_LASAT2017}         & 5/5/5 & 8.7   & 25.0 & 25  & 4.4    & 25.0  & \textbf{25}  \\
$3^{10}$                                                                                 & 29~\cite{Konishi2019arXiv}          & 5/5/5 & 10.5  & 26.6 & 26  & 4.0    & 26.6  & \textbf{26}  \\
$3^{11}$                                                                                 & 32~\cite{Konishi2019arXiv}          & 5/5/5 & 13.3  & 27.6 & 27  & 6.3    & 27.6  & \textbf{27}  \\
$3^{12}$                                                                                 & 34~\cite{Konishi2019arXiv}          & 5/5/5 & 16.5  & 28.2 & 28  & 8.5    & 28.2  & \textbf{28}  \\
$3^{13}$                                                                                 & 36~\cite{Konishi2019arXiv}          & 5/5/5 & 16.3  & 29.6 & 29  & 6.5    & 29.6  & \textbf{29} \\ \hline
\end{tabular}

\end{table}
The third experiment campaign aims to show  
how close the locating arrays obtained by the proposed approach 
are to the smallest-known arrays with respect to the number of rows. 

At present, to our knowledge, the concrete sizes of locating arrays are known 
for only a small number of problem instances. 
Table~\ref{tab:known} lists several of such $(\overline{1}, 2)$-locating arrays, where 
the problem instance, known array size, and source reference are shown for each array 
in the three left columns.
We applied the proposed approach to these problems and obtained the 
results summarized in the remaining columns in the same table, as in Tables~\ref{tbl:result2}
and \ref{tbl:result3}. 
Since these problems are relatively small, the algorithm 
was always successful in completing its execution within one hour, 
except for the largest instance. 
Even for that instance, the algorithm was able to produce 
a locating array in every run of the program. 

Numbers in bold font  represent the 
smaller array size between the known locating array or the one 
obtained in the experiments. 
The known array sizes underlined 
are proved to be optimal (minimum). 
Such optimal arrays have been known only for small problems. 
The results show that the arrays obtained by applying our algorithm to these small problems 
were also optimal or very close to optimal. 

For larger instances, our algorithm was able to find locating arrays 
that are considerably smaller than known ones. 
The size reduction achieved is substantial, particularly
for the first two instances which have arisen from real-world applications~\cite{Aldaco:2015,Compton2016}. 
The $(\overline{1}, 2)$-locating array of size~421 for the first instance~\cite{Aldaco:2015} 
was ``crafted from tools for covering arrays, using simple heuristics''\cite{colbourn2016}.
The array we obtained using the single algorithm is smaller than this previously known array 
by more than 30 percent. 
For the second instance, the locating array we found is also approximately 30 percent smaller 
than the previously known one. 

\section{Related work}\label{sec:realtedWork}

Colbourn and McClary proposed locating arrays in~\cite{colbourn_locatingarray2008}.
They suggested a few possible applications of locating arrays, 
including combinatorial interaction testing, group testing, etc. 
Locating arrays were used for screening interacting factors in 
wireless network simulator and testbed in~\cite{Aldaco:2015,7562157}. 
Applications to combintatorial interacting testing were discussed in~\cite{7528941}.

Mathematical aspects of locating arrays have been explored in~\cite{tang_optimalitylocatingarray_2012,Shi2012,Spread2016}. 
These studies assume that the number of values that a parameter can take 
is the same for all parameters. 
In~\cite{tang_optimalitylocatingarray_2012}, Tang, Colbourn, and Yin proved 
a lower bound on the size of locating arrays. This bound was used in our proposed algorithm, as described in Section~\ref{sec:algorithm}. 
In~\cite{Shi2012}, an infinite but sporadic series of minimum $(\overline{2}, t)$-locating arrays are provided. 
In~\cite{Spread2016}, the constructions of the minimum $(1, 1)$-, $({\overline 1}, 1)$-, $(1, {\overline 1})$-, and $({\overline 1}, {\overline 1})$-locating arrays are presented. 
Mathematical constructions of $(1, \overline{2}$)-locating arrays 
are proposed in~\cite{colbourn2016}.

Computational constructions can naturally handle the case where the domain sizes of 
parameters are different for different parameters. 
In \cite{nagamoto_locatingpairwisetesting2014}, 
a ``one-test-at-a-time'' greedy heuristic
algorithm is proposed to construct $(1, 2)$-locating arrays.
In our previous work~\cite{konishi_LASAT2017,Konishi2019arXiv}, we proposed 
an approach that uses a Constraint Satisfaction Problem (CSP) solver 
to find locating arrays. 
The arrays obtained by using the CSP-based approach were compared 
with those obtained by the simulation annealing-based algorithm in 
Section~\ref{sec:results}. 
Recently, techniques based on \textit{resampling}~\cite{Moser:2010:CPG:1667053.1667060} 
have been investigated for locating array construction.  
Studies in this line include \cite{10.1007/978-3-319-94667-2_29,Colbourn2018}. 

In contrast to the small amount of research on computational construction of locating arrays, 
many approaches have been applied to the problem of generating covering arrays. 
These approaches include simulated annealing~\cite{garvin_metaheuristic2011,776822}, 
as well as 
greedy heuristics~\cite{Bryce:2005:FGM:1062455.1062495,cohen_AETG1997}, 
tabu search~\cite{lin_TCA2015,GECCO17}, 
genetic algorithms~\cite{Shiba:2004}, 
particle swarm~\cite{6919298,AHMED201720}, 
and others~\cite{ALSEWARI2012553,AHMED201513}.
Our proposed algorithm for generating 
locating arrays shares some similarities 
with the simulated annealing-based covering array 
constructions but 
is different in many aspects, 
not only in technical details but also in fundamental 
algorithm designs. 
For example, the cost function of simulated annealing 
and neighbor selection strategies are completely different.  

\section{Conclusion}\label{sec:conclusion}

This paper proposed a simulation annealing-based algorithm to construct $(\overline{1}, t)$-locating arrays. 
In the algorithm, simulation annealing is used to 
find a locating array of a given size. 
By repeatedly running simulated annealing with the size being systematically varied, 
the algorithm eventually yields a locating array of a small size. 
Experimental results showed that when $t = 2$, the proposed algorithm 
scales to large-sized problems and is able to produce 
locating arrays that are substantially smaller than those previously known. 
The results also showed that for small problems, 
the $(\overline{1}, 2)$-locating arrays 
produced by the algorithm were minimum or close to minimum. 

Future research includes several directions. 
For example, improving the scalability of the algorithm 
deserves further study, since as the experimental 
results shows, the current algorithm is often 
unable to produce a locating array when the strength $t$ of the array exceeds two. 
More fine-grained tuning of parameter values of simulated annealing may also require further research. 
Machine learning techniques might be useful for this problem. Another possible direction is to adapt the proposed 
algorithm to construction of other related mathematical 
objects that can be used for fault localization. 
Examples of such mathematical objects include  
\emph{detecting arrays}~\cite{colbourn_locatingarray2008}, 
\emph{error locating arrays}~\cite{MartinezELA2010}, 
and \emph{constrained locating arrays}~\cite{DBLP:journals/corr/abs-1801-06041}.


\end{document}